\documentclass[epsf,usegraphicx]{mn2e}

\title[RATS: Overview and first results]{RApid Temporal Survey
- RATS I: Overview and first results}

\author[Ramsay \& Hakala]
{Gavin Ramsay$^{1}$, Pasi Hakala$^{2}$\\
$^{1}$Mullard Space Science Lab, University College London,
Holmbury St. Mary, Dorking, Surrey, RH5 6NT, UK\\
$^{2}$Observatory, University of Helsinki, PO Box 14, 
FIN-00014 University of Helsinki, Finland}

\begin{document}
\outer\def\gtae {$\buildrel {\lower3pt\hbox{$>$}} \over 
{\lower2pt\hbox{$\sim$}} $}
\outer\def\ltae {$\buildrel {\lower3pt\hbox{$<$}} \over 
{\lower2pt\hbox{$\sim$}} $}
\newcommand{\ergscm} {ergs s$^{-1}$ cm$^{-2}$}
\newcommand{\ergss} {ergs s$^{-1}$}
\newcommand{\ergsd} {ergs s$^{-1}$ $d^{2}_{100}$}
\newcommand{\pcmsq} {cm$^{-2}$}
\newcommand{\ros} {\sl ROSAT}
\newcommand{\exo} {\sl EXOSAT}
\def\rchi{{${\chi}_{\nu}^{2}$}}
\newcommand{\Msun} {$M_{\odot}$}
\newcommand{\Mwd} {$M_{wd}$}
\def\Mdot{\hbox{$\dot M$}}
\def\mdot{\hbox{$\dot m$}}

\maketitle

\begin{abstract}

We present the aim and first results of the RApid Temporal Survey
(RATS) made using the Wide Field Camera on the Isaac Newton
Telescope. Our initial survey covers 3 square degrees, reaches a depth
of V$\sim$22.5 and is sensitive to variations on timescales as short
as 2 minutes: this is a new parameter space. Each field was observed
for over 2 hours in white light, with 12 fields being observed in
total.  Our initial analysis finds 46 targets which show significant
variations. Around half of these systems show quasi-sinusoidal
variations: we believe they are contact or short period binaries. We
find 4 systems which show variations on a timescale less than 1
hour. The shortest period system has a period of 374 sec. We find two
systems which show a total eclipse.  Further photometric and
spectroscopic observations are required to fully identify the nature
of these systems. We outline our future plans and objectives.

\end{abstract}


\section{Introduction}

The intensity of stellar objects can vary on a wide range of
time-scales, ranging from seconds to months to years. A large number
of projects now exist whose aim is to detect such varying sources. The
reasons for this are many, but include the search for extra-solar
planets and interacting binary stars. Most of these surveys are
sensitive to timescales longer than a day. It is only recently that
such surveys have been sensitive to shorter term timescales. For
instance, the $0.3^{\circ}\times0.3^{\circ}$ survey using the Canada
France Hawaii Telescope was sensitive to variations on timescales as
short as $\sim$8 min (Lott et al 2002). While the Faint Sky
Variability Survey has a much larger survey area, it was sensitive to
variations only as short as $\sim$24 min (Groot et al 2003,
Morales-Rueda et al 2004). In principle the SuperWasp project is
sensitive to variations on timescales as short as a few mins
(Christian et al 2004). However, they are sensitive to relatively
bright objects, $V\sim$7--15.

Why is it important that we extend the parameter search down to
periods shorter than 10 min? Recently a new class of object has been
discovered in which coherent intensity variations have been detected
on timescales of $\sim$10 min or less, with the shortest being 5.4
mins (see Cropper et al 2004 for a review). It is thought that these
systems are interacting white dwarf-white dwarf pairs which have no
accretion disc, and the observed period represents the binary orbital
period. As such, they are expected to be amongst the first sources to
be detected using {\sl LISA}, the planned gravitational wave observatory
(Nelemans, Yungleson \& Portegies Zwart 2004).

These systems are at the short period end of the period distribution
of white dwarf-white dwarf binaries, or AM CVn systems. For orbital
periods less than 80 min, the secondary (mass donating) star cannot be
a main sequence star. Further, for periods shorter than 30 min, the
secondary must be a Helium white dwarf (Rappaport, Joss \& Webbink
1982). There are around 13 of these AM CVn systems currently known.

Theoretical predictions suggest that AM CVn systems are commonplace
(eg Nelemans et al 2001). It is not clear if the discrepancy between
the observed number and predicted number of systems is due to an over
estimate in the predicted number or if many more systems await
discovery. To help resolve this question we have started a project,
the RApid Temporal Survey (RATS), whose aim is to discover new objects
which vary in a coherent manner on timescales less than $\sim$1hr. As
a significant by-product of this search we expect to discover many new
objects which vary on longer periods. This paper presents our strategy
for discovering new AM CVn systems; the positions and light curves of
those systems which we have discovered as part of our initial set of
observations; a discussion regarding the nature of these sources and a
brief discussion as to what these observations imply regarding the
population density of AM CVn systems and other objects.

\section{Observations and Reduction}

\subsection{Observational strategy}

Our observations were made over 3 nights starting on 28 Nov 2003 using
the 2.5m Isaac Newton Telescope on La Palma and the Wide Field Camera
(WFC). The WFC consists of 4 EEV 2kx4k CCDs, with each pixel
corresponding to 0.33$^{''}$ on the sky. The approximate field of view
is 33$^{'}\times33^{'}$ (with a 11$^{'}\times11^{'}$ gap missing on
one of the corners). The orientation of the instrument was such that
our field center was located near the center of CCD 4 (see Groot et al
2003 for a schematic diagram showing the orientation of the chips).

Since our goal was to obtain photometry with the highest possible time
resolution, we did not use a filter - eg they were `white light'
observations: since our observations were made near the zenith,
differential atmospheric diffraction was reduced to a
minimum. However, at the start of each observation we also obtained
$BVI$ images. Since we did not take observations of standard stars
before every field observation, the $BVI$ magnitudes reported in Table
2 are probably accurate to only $\sim$0.1 mag. Our exposure times were
30 sec in duration, with another $\sim$42 sec of readout time for the
whole array. Each field was observed for around 2-2.5 hrs. We obtained
bias frames and flat fields on each night.

\begin{table}
\begin{center}
\begin{tabular}{rrrrr}
\hline
Field & $\alpha$ & $\delta$ & $l$ & $b$ \\
\hline
1 & 02h 02m & +34$^{\circ}$ 21$^{'}$ & 139$^{\circ}$ & --26$^{\circ}$\\
2 & 02h 08m & +36$^{\circ}$ 16$^{'}$ & 140$^{\circ}$ & --24$^{\circ}$\\
3 & 03h 07m & --00$^{\circ}$ 31$^{'}$ & 179$^{\circ}$ & --48$^{\circ}$ \\
4 & 04h 11m & +19$^{\circ}$ 25$^{'}$ & 174$^{\circ}$ & --23$^{\circ}$\\
5 & 04h 51m & +18$^{\circ}$ 11$^{'}$ & 181$^{\circ}$ & --16$^{\circ}$ \\
6 & 04h 56m & +13$^{\circ}$ 05$^{'}$ & 187$^{\circ}$ & --18$^{\circ}$\\
7 & 07h 29m & +23$^{\circ}$ 25$^{'}$ & 195$^{\circ}$ & +18$^{\circ}$\\
8 & 07h 40m & +23$^{\circ}$ 50$^{'}$ & 196$^{\circ}$ & +21$^{\circ}$ \\
9 & 08h 06m & +15$^{\circ}$ 27$^{'}$ & 207$^{\circ}$ & +23$^{\circ}$\\
10 & 22h 57m & +34$^{\circ}$ 13$^{'}$ & 97$^{\circ}$ & --23$^{\circ}$\\
11 & 23h 04m & +34$^{\circ}$ 26$^{'}$ & 107$^{\circ}$ & --26$^{\circ}$\\
12 & 23h 10m & +34$^{\circ}$ 19$^{'}$ & 100$^{\circ}$ & --24$^{\circ}$\\
\hline
\end{tabular}
\end{center}
\label{log}
\caption{The field centers for our observations. The co-ordinates are
in J2000}
\end{table}

\subsection{Field selection}

Our aim was to select fields which were close to the galactic plane so
that there were as many stars in the field as possible, but not so
close to the plane that crowding became a serious issue. Further, we
wanted to select fields that did not contain stars brighter than
$\sim$10 mag, which might cause charge overflow on the CCDs. To make
this selection we obtained the APM
catalogue\footnote{http://www.ast.cam.ac.uk/$^{\sim}$apmcat} for a
range in galactic latitudes ($5<|b|<25$) and wrote a routine which
optimizes the field selections for the given sky coordinate range and
the shape and size of the field of view. For an observing run lasting
$n$ nights, our optimization routine finds the same number ($n$) of
fields around 4 positions in the sky that pass close to the zenith
during the night with about 2 hr spacing.

This is done using an algorithm that first places $n$ random fields in
the sky in a 5$\times$5 degree box centered around a given
pointing. Then a simulated annealing based optimization routine is
used to find a combination of $n$ non-overlapping pointing's within
this 5$\times$5 degree box. The optimization tries simultaneously to
maximize the number of $V$=15-20 mag stars and minimize the number of
brighter stars included. There is also a hard limit, so that no stars
brighter than $V$=10 mag should appear in any of the fields.  So, as a
result we will get, for each night, 4 fields per night that are
observable for around 2hr close to the zenith after each other, thus
filling all the nights optimally.

\subsection{Data Reduction}
\label{reduction}

A series of $\sim$100--120 30 sec exposures were made of each field:
auto-guiding was used so there is no large $x,y$ shift between
frames. For each chip we flat-fielded and bias subtracted each
image. The analysis (for each chip) consisted of following steps: the
first 5 frames of each field were combined in order to form a `master'
image of the field. This master image was then searched for all the
sources above a 5$\sigma$ detection limit. Next, we determined for
each of the remaining frames an accurate image shift in relation to
the master image. This was done in order to be able to use small
enough apertures for light curve extraction, thus maximizing the S/N
ratio for the photometry. In practice, this was achieved by using the
positions of a couple of bright (but not saturated) stars. After
having determined (and applied) the shifts for individual images, we
obtained quasi-differential photometry using the median brightness of
the sources in each frame as a comparison (to remove any variations in
the sky transparency) -- this was done for all the images of the
field. In total around 33000 stars were detected in our survey which
covered a total of 3 square degrees.

All the resulting light curves were analysed and those which had bad
magnitude values or errors above 0.1 mag or non-detections in any
frame, were discarded. This reduced the total number of light curves
by around 2000. All the `good' light curves were analysed using a
Lomb-Scargle period search algorithm, and those light curves which
showed a significant peak in their power spectrum were flagged for
closer inspection. The significance of the highest peak in each power
spectrum was estimated in the following manner. First, we took the
highest peak power value and mean power value for each power spectrum.
Then we defined the significance statistic of the highest peak in the
power spectrum by computing the ratio of it to the mean power
value. The final selection of `interesting' variable targets was
carried out manually. For each chip we produced a plot of apparent
magnitude against the significance statistic for all the stars. When
the overall number of fields is not too large (as in the case here),
such a plot provides a convenient way of of identifying the likely
variables in each field.

\begin{figure}
\begin{center}
\setlength{\unitlength}{1cm}
\begin{picture}(8,6.5)
\put(-2.5,-7.5){\includegraphics{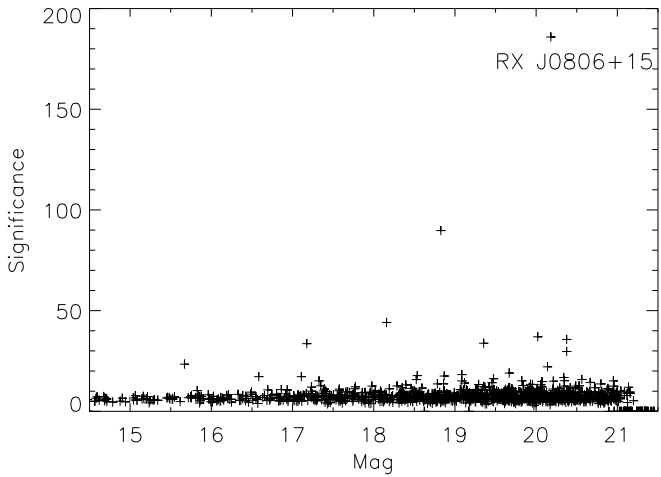}}
\end{picture}
\end{center}
\caption{The results of our routine to detect sources varying in their
intensity using the field of the
ultra-compact binary RX J0806+15 as a test case. The magnitudes have been
scaled so they are approximately that of $V$ mag and higher significance 
values imply strong variability. Our routine clearly
identifies RX J0806+15 as being highly variable. }
\label{rx0806}
\end{figure}

To test our procedure we selected the field of RX J0806+15, the 5.4
min ultra-compact binary system. Its brightness variation is highly
stable with a peak-to-peak modulation of $\sim$0.3 mag in white light
(Ramsay, Hakala \& Cropper 2002). We ran our data analysis procedure
in the manner outlined above. For each star in the field we plot the
significance of the highest peak in its power spectra (as defined
above) as a function of brightness in Figure \ref{rx0806}. RX J0806+15
was easily picked up by our routine. We are therefore confident that
we can detect sources which have coherent variability with amplitudes
of $\sim$10 percent or lower.

For those sources identified as being significantly variable, we used
{\tt autophotom} (Eaton, Draper \& Allan 2003) to obtain `optimal'
photometry for each source together with suitable comparison sources.
Whilst our procedure for identifying variable sources is relatively
simple, we have demonstrated that it will detect targets such as RX
J0806+15 with ease. However, it is probable that additional variable
sources still await detection using a more sophisticated detection
routine.

\subsection{Source positions}

To determine accurate positions for our variable sources, we compared
our fields with the Digital Sky Survey images and the Hubble Guide
Star Catalogue. We typically used 8 reference stars for each chip. We
then compared the $x,y$ position of each reference stars in our images
and used {\tt astrom} (Wallace \& Gray 2002) to obtain J2000
co-ordinates for each of our sources. The residuals to the fits were
typically $\sim0.5^{''}$. We searched for known sources in the SIMBAD
database using a aperture of radius 10$^{''}$ centered on our source
positions: there were no known sources at these positions. All our
sources are therefore newly discovered variable sources.

\section{Initial results}
\label{results}

Out of the 33000 stellar objects in our survey, we have identified 46
which have showed evidence for significant variability. Their
position, $V$ mag and colour and the characteristics of their light
curves are shown in Table \ref{table_results}.

\begin{figure}
\begin{center}
\setlength{\unitlength}{1cm}
\begin{picture}(8,10.5)
\put(-0.5,4){\includegraphics{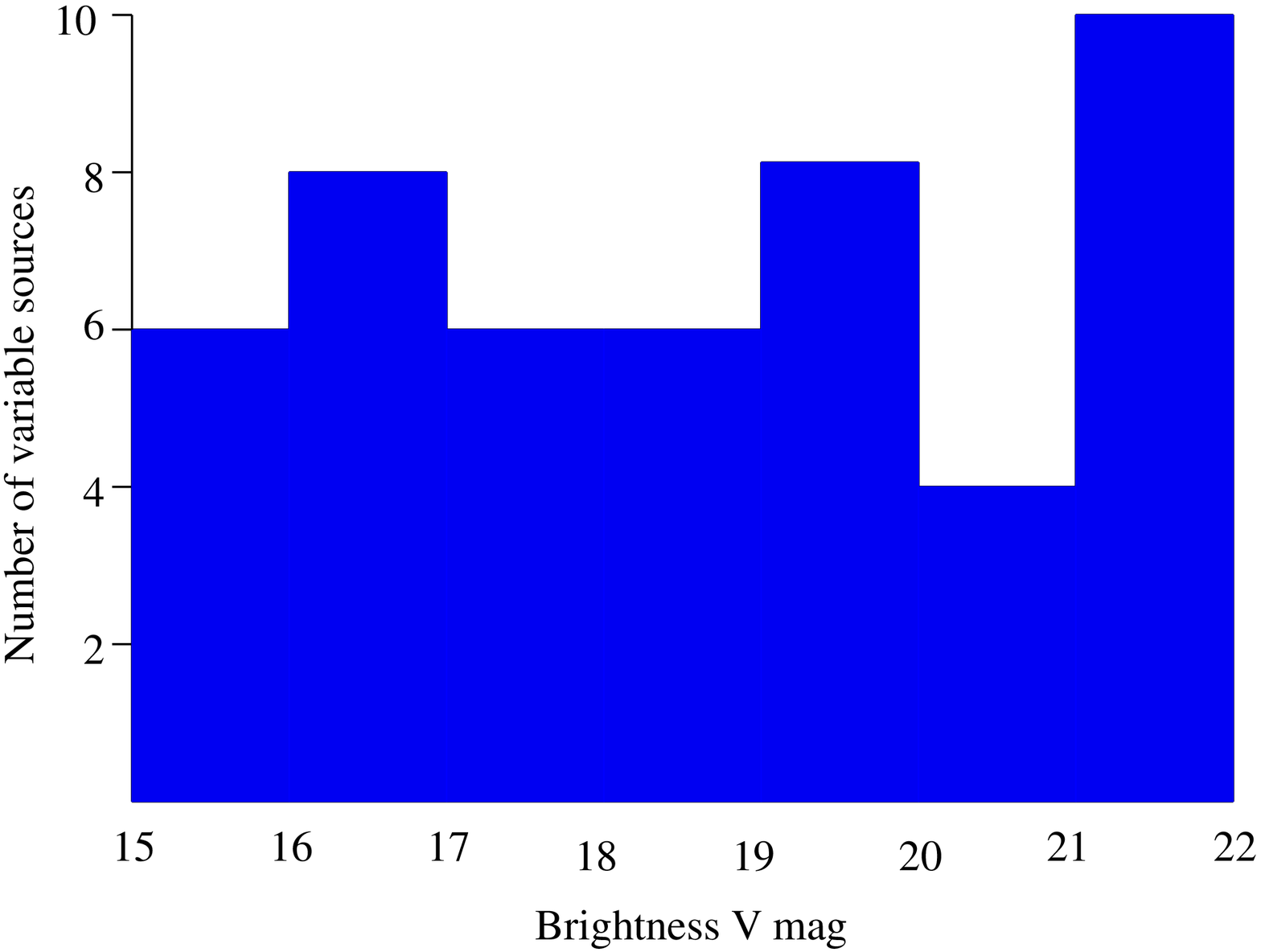}}
\put(-0.5,-0.5){\includegraphics{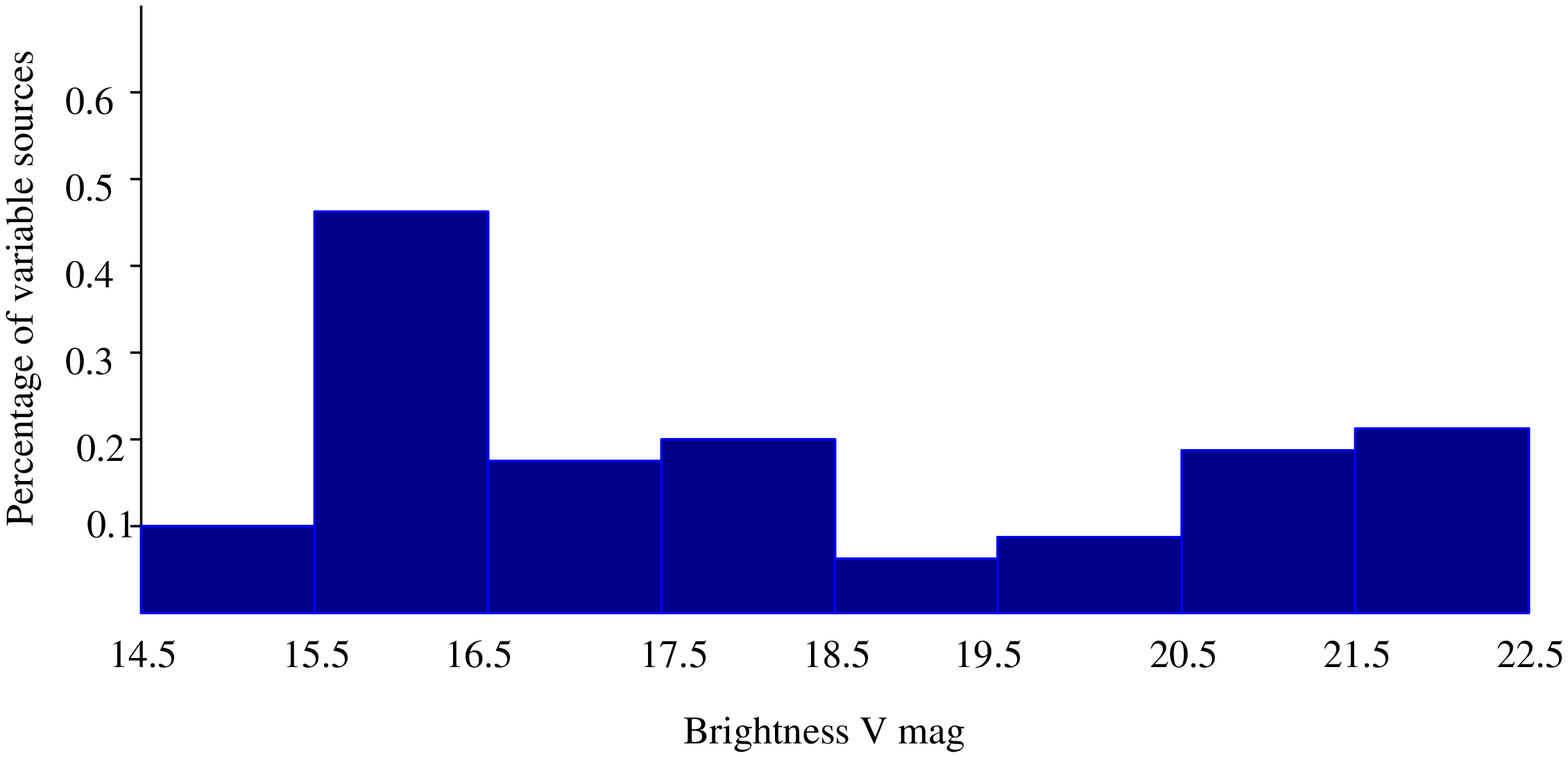}}
\end{picture}
\end{center}
\caption{Top panel: The number of variable source discovered in our
survey as a function of magnitude. Bottom panel: The percentage of
variable sources as a function of the number of stars in a magnitude range.}
\label{number}
\end{figure}

We show in Figure \ref{number} the actual number of sources which
varied as a function of apparent brightness and also the fraction of
sources which were found to be variable as a function of
brightness. We find variable sources in the range $15<V<22$, with an
approximately uniform spread in the number of sources as a function of
apparent brightness. Although, the number of sources in the bins are
relatively small, we note that the fraction of sources which are
variable is not uniform, showing a peak in the bin $15.5<V<16.5$. We
discuss one possible reason for this in \S 4.

The light curves of our 46 variable sources are shown in Figures 3--5.
They can be split up into 5 broad categories: those showing
sinusoidal-like variations; coherent variations; eclipse-like
behaviour; flaring; and more irregular behaviour. We now go onto
discuss various objects in more detail.

\begin{figure*}
\begin{center}
\setlength{\unitlength}{1cm}
\begin{picture}(16,23)
\put(0,0){\includegraphics{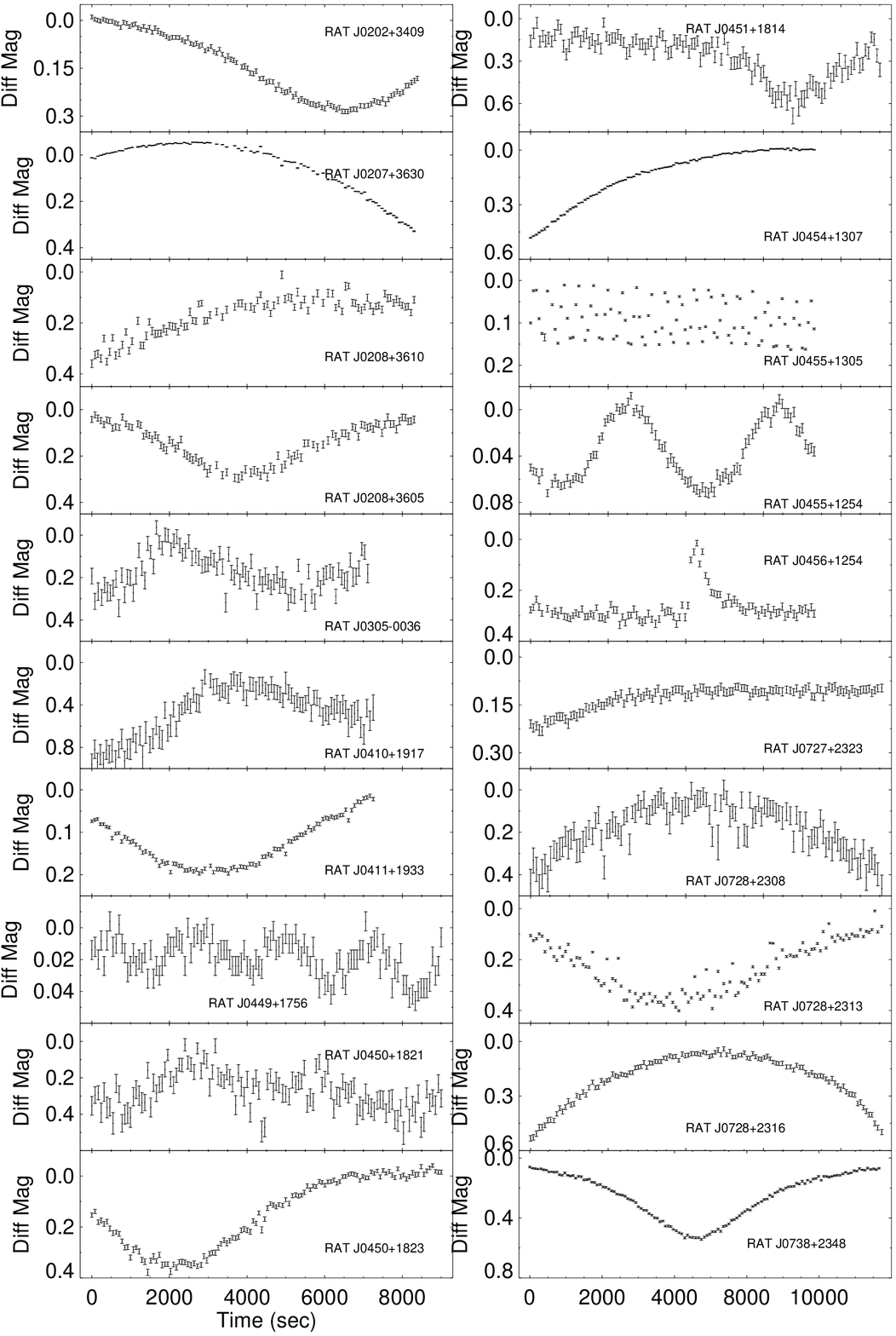}}
\end{picture}
\end{center}
\caption{Light curves of all interesting sources.}
\label{curves1}
\end{figure*}

\begin{figure*}
\begin{center}
\setlength{\unitlength}{1cm}
\begin{picture}(16,23)
\put(0,0){\includegraphics{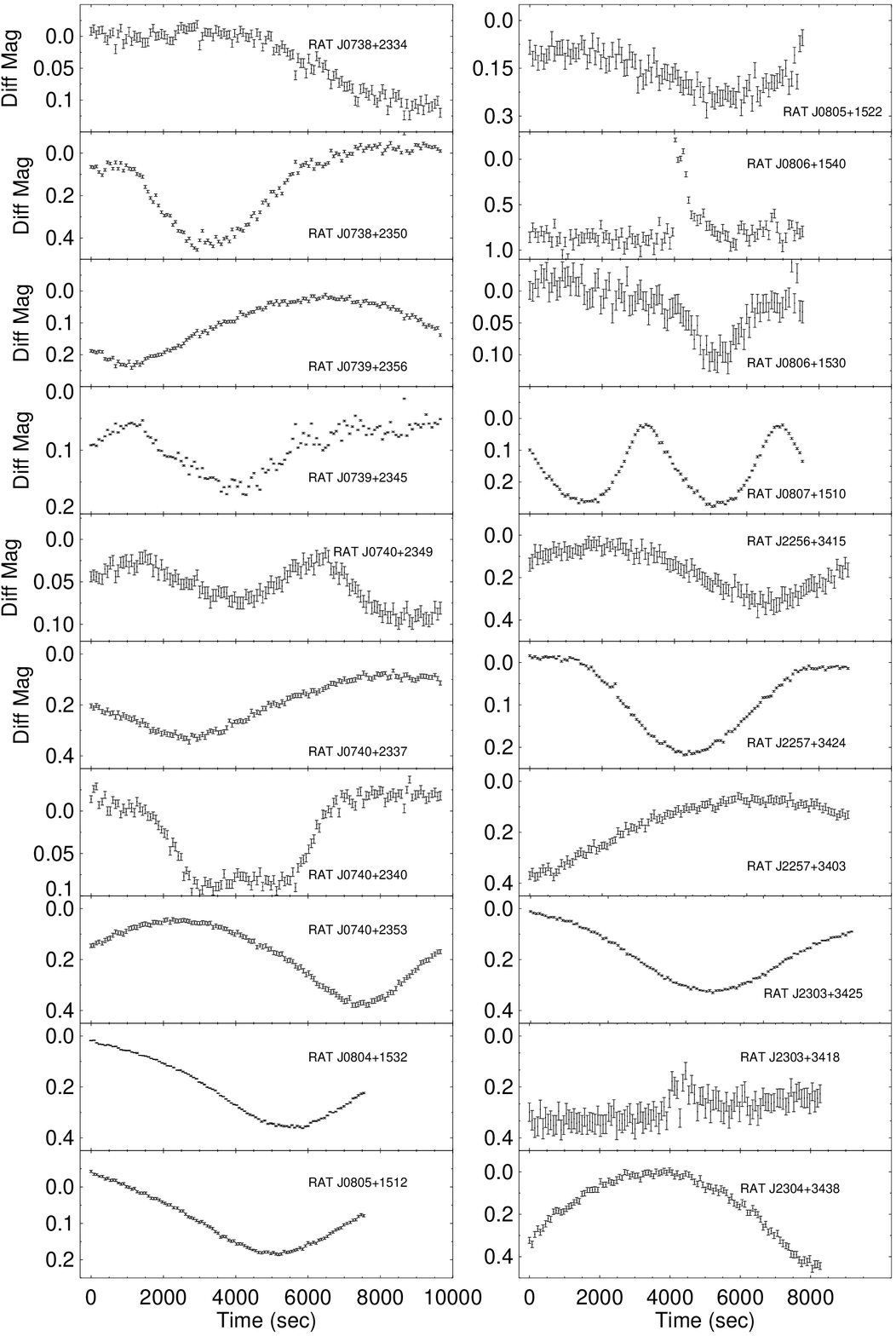}}
\end{picture}
\end{center}
\caption{Light curves of all interesting sources (cont).}
\label{curves2}
\end{figure*}

\begin{figure*}
\begin{center}
\setlength{\unitlength}{1cm}
\begin{picture}(16,8)
\put(0,-15){\includegraphics{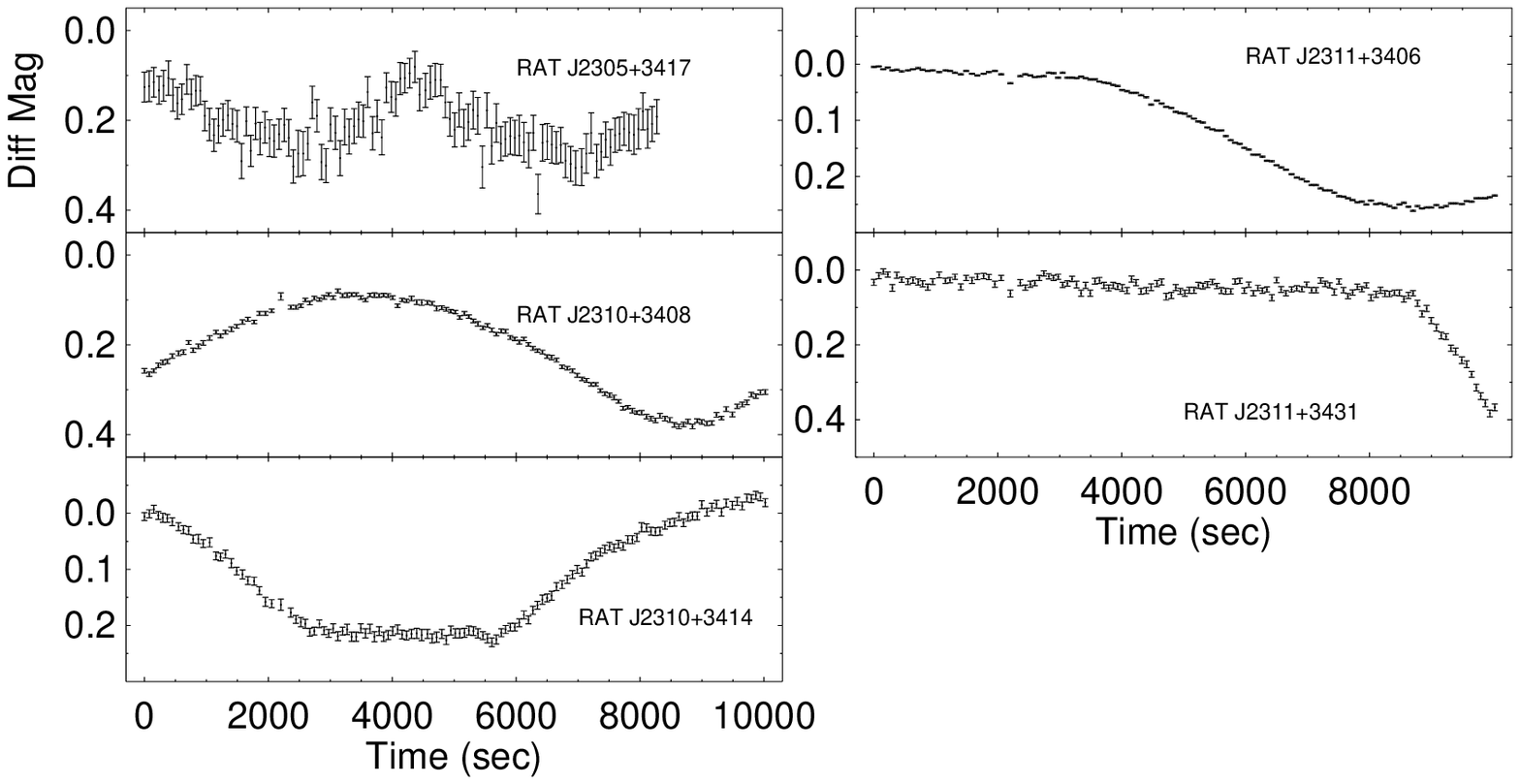}}
\end{picture}
\end{center}
\caption{Light curves of all interesting sources (cont).}
\label{curves3}
\end{figure*}


\begin{table*}
\begin{center}
\begin{tabular}{lrrrrrrrr}
\hline
Field & RA  & Dec  & Period & $\Delta$ M & Feature & $V$ & $B-V$ &
$V-I$ \\
\hline
1 & 02 02 22.3 & +34 09 09.8 & :3.5hr & 0.3m & sin  &   17.3 & 
0.3 & 0.7\\
2 & 02 07 31.4 & +36 30 30.5 & & :0.4m & sin & 15.7 & 0.3&  0.7 \\
2 &  02 08 03.9&  +36 10 16.7 & & :0.3m & rise     & 
19.6 & 0.5&  0.7\\
2$^{*}$ & 02 08 40.4 & +36 05 31.1 & 4hr? & :0.3 & sin   & 18.8 & 0.4& 0.8\\ 
3 & 03 05 56.8&  --00 36 16.5  & 100min? & :0.3m  & sin &   20.2&
0.3 &  0.6\\
4 & 04 10 18.2 & +19 17 12.2 & & :0.6m&  :sin & 21.3 & 0.6 & 0.8\\
4 & 04 11 20.1 & +19 33 59.8 & & 0.2m & sin  & 16.8 &  0.6 &  1.0\\
5 & 04 49 52.3& +17 56 37.8 &40/80$^{'}$ & ~0.04m& regvar & offchip &
& \\
5 & 04 50  02.7 & +18 21 07.4 & &  :0.4m & var  & 21.1  & 0.6 & 0.9\\
5 & 04 50 15.6&  +18 23 44.7& :3hr & :0.4m& sin & 18.4& 0.7 & 1.1\\
5 & 04 51 21.9&  +18 14 11.0& & :0.6m & dip. & 21.6 & & 2.2\\ 
6 & 04 54 29.4 & +13 07 22.8 & & 0.5m & sin &  16.4&  
0.6&  1.1\\
6  & 04 55 15.2 & +13 05 29.7 & 374sec & 0.15m& sin& 17.2 & 0.0
& 0.4 \\  
6 &  04 55 16.5 & +12 54 10.5 & 1.1hr & 0.08m & sin & 16.0&  
0.4&  0.8\\
6 &  04 56 06.9&  +12 54 15.4 & & 0.3m & flare &
21.1 & 1.6 &  2.9\\
7 & 07 27 55.5 &+23 23 24.8 & & 0.1m & rise & 19.4 & 0.4 & 0.8\\
7 & 07 28 31.9 & +23 08 52.4 & & 0.4m & sin& 21.5  & 0.6& 0.8\\ 
7 & 07 28 49.2 & +23 13 38.2 & & 0.3m & sin& 15.7  & 0.4&  0.8\\
7 & 07 28 53.5 & +23 16 54.0 & & 0.5m & sin& 19.9  & 0.9& 1.1\\
8 & 07 38 51.3 & +23 48 08.9 &  & :0.5m & sin &  16.5 &
0.7 &  1.0\\
8 & 07 38 57.1 & +23 34 25.3 &  & :0.1m & :sin& 18.2  & 0.6 & 
0.8\\
8 & 07 38 59.2 & +23 50 37.8 &  & :0.4m & dip & 16.1 & 0.6 & 0.9\\
8 & 07 39 24.8 & +23 56 26.1 & :3hr & :0.2m & sin & 17.5 & 0.7
& 0.9\\
8 & 07 39 31.5 & +23 45 37.7 & & :0.1m &  dip & 16.0  & 1.4
& 0.8\\
8 & 07 40 06.1  & +23 49 28.6 & & :0.1m & var & 17.8 &0.2 &0.5\\  
8 & 07 40 16.7 & +23 37 53.4 & :3hr & :0.25m & sin & 18.4&
0.5& 1.0\\
8 & 07 40 17.1 & +23 40 09.0 &  &  ~0.1m & ecl  &18.1& 0.7 & 1.1\\
8 & 07 40 24.4 & +23 53 19.5 & 2--3hr? & :0.4m & sin &  18.8& 0.6&
1.6\\
9 & 08 04 50.7 & +15 32 11.7 & 3.1hr& 0.4m &  sin &            
15.6&  0.5&  0.8\\
9 & 08 05 16.2 & +15 12 58.5 & :2.8hr & 0.2m & sin  &          
16.1&  0.6&  0.8\\
9 & 08 05 40.1 & +15 22 59.0 & & 0.2m & :sin & 20.4 & 0.6 &  1.0\\
9 & 08 06 29.1 & +15 40 45.8 & & 1m & flare  & & & $I\sim$22.0\\
9 & 08 06 44.5 & +15 30 00.9 & :2.5hr & 0.1m & dip  &    
21.1  & 1.34  & 2.8\\
9 & 08 07 00.5 & +15 10 58.2 & 1hr& 0.3m  &  sin  &      
15.4 & 0.00 &  0.2\\
10 & 22 56 54.2 & +34 15 38.7 & 4hr? & :0.3m &  & 21.2 & 0.8 & 1.6 \\
10 & 22 57 20.2 & +34 24 30.2 & & :0.2m & gecl?   &   16.0 & 0.7 & 1.1\\
10 & 22 57 45.2 & +34 03 27.3 & 6hr? & :0.3m & sin  & 19.9 & 1.0 & 0.9\\  
11 & 23 03 04.9 & +34 25 29.6 &  200min? & 0.3m & sin & 16.5 & 0.6& 1.0\\
11 & 23 03 19.3 &  +34 18 45.8 &  & :0.2m & var & 22.2 & - & 2.9\\
11 & 23 04 28.6 &  +34 38 54.2 &  $>$4hr & :0.5m & sin & 20.14 & 1.3 & 1.9\\
11 & 23 05 07.9 & +34 17 23.7 & 75min? & 0.2m & sin & 20.9 & 0.3 & 0.5\\
12  & 23 09 58.4 & +34 35 53.9 & & 0.4m   & var    & 21.31 & 
0.7 & 1.0 \\
12  & 23 10 25.7 & +34  08 44.6 & 2.9hr & 0.3m  & sin & 17.7&
0.7&  1.0\\
12  & 23 10 38.6 & +34 14 33.1 & 3hr & 0.25m  & ecl & 17.9 &
0.4 & 0.6\\
12  & 23 11 15.6 & +34  06 44.8 & :4hr & 0.3m & sin  & 15.8&
0.4 & 0.8 \\
12  & 23 11 24.7 & +34 31 17.7 & & 0.4m & into ecl?  & 
19.8 & 1.1&  1.8\\
\hline
\end{tabular}
\end{center}
\label{results}
\caption{Those sources discovered in the first reduction of our
INT/WFC survey data. We quote the period of any source showing
periodic behaviour, together with the change in white light magnitude
and the main characteristic of the light curve (sin: sinusodial in
shape; ecl: eclipse; shortp: short period variability). $^{*}$RAT
J0208+3605 is $\sim1^{''}$ distant from another source from which it
was not possible to fully resolve: the quoted magnitudes are the
combined.}
\end{table*}

\label{table_results}

\subsection{Objects varying on periods of less than 1hour}

The main goal of this project is to discover sources which vary in
their intensity in a coherent manner on timescales of $\sim$1 hour or
less.  We have discovered one source with a coherent period of 374
sec, 2 systems showing sinusoidal variations on timescales of $\sim$1
hour, one which shows a variation on 40 or 80 mins and one with a
possible variation on a period of 75 mins.

\begin{figure}
\begin{center}
\setlength{\unitlength}{1cm}
\begin{picture}(8,4)
\put(-1.5,-6){\includegraphics{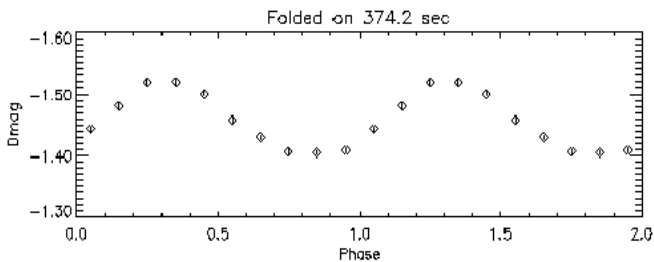}}
\end{picture}
\end{center}
\caption{White light data of RAT J0455+1305
folded on the best fit period of 374 sec.}
\label{rat0455_lc}
\end{figure}

The system with the shortest period was RAT J0455+1305 which was found
to vary on a period of 374 sec and an amplitude of 0.13 mag.  The
source is detected in the USNO-B1 Catalogue, but not in the {\sl
ROSAT} all-sky survey. The folded light curve appears sinusoidal
(Figure \ref{rat0455_lc}).  The shape and amplitude are virtually
identical to the discless ultra-compact binaries mentioned in \S 1. On
the other hand pulsating white dwarfs and sub-dwarf pulsators (EC14026
stars) also show such coherent modulations. Dedicated observations of
this object will be presented in a future paper.

The two systems which show sinusoidal variations on periods close to
1hr, RAT J0455+1254 and RAT J0807+1507, have amplitudes of 0.06 and
0.25 mag respectively. The latter has a light curve which shows
clearly peaked maxima. If their periods are indeed close to 1 hr, then
this would rule out them being W UMa systems - contact binaries which
have orbital periods in the range 0.2--1.5 days.  

RAT J0449+1756 shows clear variations with an amplitude of $\sim$0.04
mag. It is unclear whether there is a modulation on a period of 40 or
80 min. Unfortunately this object was just off the chip edge when we
made the $BVI$ images. It was, however, detected in the 2MASS survey:
it shows $JHK$ colours which are blue. Its blue colour and its period
are similar to that of the AM CVn systems: in this case, it would be
expected to show strong Helium lines in emission in its optical
spectrum.

\subsection{Objects showing eclipse-like features}

There are three sources which show eclipse-like behaviour: RAT
J0740+2340 (a total eclipse of depth 0.1 mag), RAT J2310+3414 (an
eclipse-like feature of depth $\sim$0.2 mag) and RAT J2257+3424 (a
partial eclipse with depth 0.2 mag). Each system shows red optical
colours out of eclipse (cf Table 2) and were all detected in the 2MASS
survey\footnote{http://pegasus.phast.umass.edu}: RAT J0740+2340 has
colours of a M4-5 dwarf star while RAT J2257+3424 is consistent with
an early K dwarf star. In the case of RAT J0740+2340 the white light
and $BVI$ images show a slight elongation of the source implying the
colours for this source may be contaminated by a close `companion'
star. RAT J2310+3414 shows infrared colours, (H-K)=0.52, (J-H)=-0.08,
which are not consistent with that of main sequence stars.

The eclipse of RAT J0740+2340 takes $\sim$20 min to descend into and
out of eclipse. The total duration of the eclipse is $\sim$65 min
(half eclipse depth). The rather low eclipse depth rules out it being
a cataclysmic variable which would show a much greater eclipse
depth. RAT J2257+3424 maybe an Algol system in which we have detected
the secondary eclipse.

As the shape of the eclipse profile of RAT J0740+2340 closely
resembles one produced by a planetary transit, we have studied it's
properties further. Assuming the eclipse is produced by a transiting
planet, we can work out some preliminary parameters for the
system. Firstly, the 0.1m depth of the eclipse implies that the radius
of the planet has to be $\sim$0.3 of the stellar radius. Secondly, the
ratio of ingress/egress lengths to the duration of the total part of
the eclipse implies that the transit occurs at an apparent stellar
latitude of about 30$^{\circ}$.
        
Some preliminary spectra of RAT J0740+2340 show that the star is of
very late spectral type. Assuming a M4-6V spectral type, this would
imply a stellar radius of $\sim$0.3 R$_{\sun}$ and planetary radius of
about 70000 km, i.e.  the same as Jupiter. Using the stellar radius
and the implied transit latitude together with the FWHM eclipse
duration yields an orbital velocity of 90 km/sec, which in turn leads
to an orbit of roughly 3 days. This is similar to that observed in a
handful of systems (eg Mazeh, Zucker \& Pont 2005). However, further
observations are required to confirm the orbital period of this
object.

\subsection{Objects showing longer term variations}

Approximately half of our newly discovered sources show sinusoidal or
quasi-sinusoidal intensity variations with periods greater than
$\sim$1hr.  Sources which exhibit these characteristics include
pulsating stars such as $\delta$ Scuti stars.  and the W UMa and Algol
binary systems.

$\delta$ Scuti stars show intensity variations on timescales between
30 min and $\sim$8 hrs, with amplitude variations less than $\sim$1
mag. However, these systems tend to show asymmetric light curves, with
the rise to maximum taking a shorter time than the descent from
maximum.  We have discovered one candidate $\delta$ Scuti system: RAT
J0305--0036 shows an asymmetric light curve; a period of $\sim$2 hrs
and an amplitude of $\sim$0.3 mag.

W UMa systems are well known systems, being thought to account for 1
in every 500 stars (Rucinski 2002) and show orbital periods between
4hrs -- 1.5 days. Other systems include binary systems in which the
mass losing star is irradiated by the mass accreting star: this
results in the companion star being apparently brighter when it is
viewed face-on compared to when we observe it from the rear. In recent
years it has become clear that many sub-dwarf B stars (sdB) are in
such binary systems: these stars are Extreme Horizontal Branch stars
which have Helium cores and only a thin layer of Hydrogen on their
surface (eg Maxted et al 2002). The companion star in these systems
could be main sequence stars, white dwarfs or even brown dwarf
stars. Those systems with white dwarfs have potentially wide
astrophysically significance since they maybe SN 1a progenitors
(Maxted, Marsh \& North 2000). We expect that many of our newly
discovered sources will be contact and non-contact binaries.

\subsection{Flare sources}

We detected two flare like objects in our survey. RAT J0456+1254
showed a 0.25 mag flare while RAT J0806+1540 showed a $\sim$1 mag
flare.  RAT J0456 was detected in all three $BVI$ bands while RAT
J0806 was detected only in the $I$ band. The former has colours
consistent with a main sequence M5/M6 star (Pickles \& van der Kruit
1990), which using the absolute magnitude relation in Allen (1973)
gives a distance of $\sim$725 pc for a M5V star and $\sim$380pc for a
M6V star. RAT J0806+1540 is therefore likely to be an even later type
dwarf star or more distant than RAT J0456. M dwarf stars are known to
show prominent flare behaviour.

\subsection{Minor Planets}

As expected a number of minor planets were detected: a total of 16
were identified. Their positions were determined for 3 epochs using
the procedure in \S \ref{reduction} : these were passed to our minor
planet colleagues based at the University of Helsinki. They determined
if there were any known minor planets at positions at these epochs - 5
were not previously known.

To accurately determine the orbit of a minor planet, and therefore
give rise to a designation for that system, positions have to be
determined on several nights. Unfortunately, since the systems were
not detected until after the observing run was over, we did not get
these positions. For future surveys we will run our source detection
software for each field in near real-time. This will allow any new
minor planet to be followed up on a subsequent night.

\section{Discussion \& Summary} 

We have presented the first results of our RApid Temporal Survey which
covered 12 INT/WFC fields, giving a total sky coverage of
approximately 3 square degrees. Our survey explores a new parameter
space - namely searching for objects which show intensity variations
on timescales as short as a few mins.  Using our relatively simple
algorithm to detect such sources we have discovered 46 sources which
showed variability on timescales between 374 sec and 4--5 hrs. For
sources which vary on periods longer than $\sim$8 hrs, it is probable
that we would not have identified them as such in our short observing
window. To determine the nature of the new variable sources, plans are
in place to obtain followup photometry and spectroscopy.

We estimate that our white light images reached a depth corresponding
to an equivalent V$\sim$22.5 (the exact depth depends on the colour of
the source). The distance to the edge of the galaxy on the far side of
the galactic center in the plane is $\sim$22.5kpc.  For a galactic
latitude of 20$^{\circ}$ (typical of fields observed in our initial
survey) we can detect every star with a spectral type earlier than
$\sim$K2 out to the edge of the galaxy.  However, the scale height of
the galaxy is of the order of 200pc. This implies that for stars
within 3 scale heights of the galaxy we can detect every star as faint
as $\sim$M5.

We noted in \S \ref{results} that there was an peak in the percentage
of objects which were variable in the range 15.5$<V<16.5$. If we
assume that at a distance of 3 scale heights the number of additional
sources which we detect is negligible then that implies a distance of
1.7kpc for galactic latitude of 20$^{\circ}$. For $V$=16 and a
distance of 1.7kpc this implies a depth of $M_{V}$=6. We note that W
UMa systems have a range in absolute magnitude of $M_{V}$=2--6 with
systems with the shortest orbital periods (0.2 days) being
intrinsically fainter than longer period systems. We suggest that if
the peak in the distribution of percentage of objects which are
variable is real (to be confirmed in the extension of the survey) then
it is due to W UMa systems: after $V\sim16$ there are no more of these
systems to discover. This is consistent with the suggestion that a
significant fraction of our newly discovered systems are W UMa
systems.

One of the main science goals of this project is to search for AM CVn
systems. Although this paper presents only our initial findings, we
have found only one or two (at most) systems which could, with further
observations, be classed as candidate AM CVn systems. How many were
expected from population synthesis models?  Nelemans et al (2001)
predicted a space density of 2.1$\times10^{-4}$ pc$^{-3}$ which was
revised down-wards by around 60 percent by Nelemans, Yungelson \&
Portegies Zwart (2004). The distance that we can detect such systems
is dependent on their absolute magnitude which in turn is dependent on
their orbital period. For longer period systems such as GP Com,
$M_{V}\sim$12, implying we could detect them out to a distance of 1.6
kpc (we expect to detect systems with shorter orbital periods out to
greater distance). If we take a uniform distribution of AM CVn
systems, then in a 3 square degree survey the above space density
predictions imply around 150 systems (with an uncertainty of at least
2). However, Nelemans et al (2004) made an investigation regarding
selection effects and found that for an optical limit of $V=20$ only
approximately 1 percent of systems with orbital periods less than 25
mins would be detected as an AM CVn system in the optical band,
implying only 1--2 systems in our 3 square square degree survey. Since
our survey extends to a fainter optical limit and work is needed to
compare the overall space density at different orbital periods, it is
too early to make definitive conclusions regarding the comparison of
theoretical predictions and observed space density estimates.

In conclusion, we have found by exploring a new parameter space around
46 new objects which have not before been identified as variable
sources.  How does this discovery rate compare with other surveys?
Mukadam et al (2004), for instance, found 35 pulsating white dwarfs in
125 nights of observing time. We have discovered one pulsating star in
3 nights, giving a similar discovery rate. However, we have found many
new objects in addition to this one very short period system. More
observational data is needed to confirm their nature: this work is in
hand, and our results will be presented in a followup paper.

\section{Acknowledgments}

PJH is an Academy of Finland research fellow.  Observations were made
using the Isaac Newton Telescope Wide Field Camera on La Palma. We
gratefully acknowledge the support of the observatory staff.  We thank
Mikael Granvik for following up our minor planet detections and Mark
Cropper for useful discussions. This research has made use of the
SIMBAD database, operated at CDS, Strasbourg, France.

{}

\end{document}